\documentclass[a4paper,12pt]{article}
\usepackage[utf8x]{inputenc}
\usepackage{graphicx}
\usepackage{amssymb,amsmath}
\usepackage{textcomp}
\def\pacs#1{\vspace{10pt} \hspace{0.33cm} \rm PACS numbers: #1 \par \vspace{10pt}}
\usepackage{ctable}
\usepackage{graphicx}
\usepackage{subfigure}
\usepackage{placeins}
\usepackage[hyperindex=true,breaklinks]{hyperref}
\usepackage[english]{babel}
\usepackage{subfigure}
\usepackage{booktabs}
\usepackage{multicol}

\begin{document}

\title{Systematic analysis of $p_T$-distributions in $p+p$ collisions.}
\author{I. Sena and A. Deppman}
\date{Instituto de Física, Universidade de São Paulo - IFUSP, Rua do Matão, Travessa R 187, 05508-900 São Paulo-SP, Brazil}

\maketitle


\begin{abstract} 
A systematic analysis of transverse momentum distribution of hadrons produced in ultra-relativistic $p+p$ collisions is presented. We investigate the effective temperature and the entropic parameter from the non-extensive thermodynamic theory of strong interaction. We conclude that the existence of a limiting effective temperature and of a limiting entropic parameter is in accordance with experimental data.
\end{abstract}

\section{Introduction}

General aspects of strong interactions up to center-of-mass energies $\sqrt{s}\sim$ 10~GeV are well understood in terms of a self-consistent theory based on the Boltzmann-Gibbs statistics \cite{Hagedorn}. Hagedorn's theory establishes a connection between the mass spectrum of highly excited hadrons and the density of states for fireballs, and provides correct descriptions for transverse momentum distributions and multiplicities of secondaries.

For $\sqrt{s}\geq$ 10 GeV, however, theoretical and experimental results diverge. A generalized formalism was proposed taking into account the non-extensive thermodynamics~\cite{Bediaga, Beck}. This generalization recovered the agreement between theory and experiment.

Recently it was shown~\cite{Deppman} that a self-consistent theory for hadronic systems based on the non-extensive thermodynamics exists if, for $x \rightarrow \infty$,
\begin{eqnarray}
\rho(x) \rightarrow \gamma x^{-5/2} e_q^{\beta_o x}
\end{eqnarray}
and
\begin{eqnarray}
\sigma(x) \rightarrow b x^{a} e_q^{\beta_o x}\,,
\end{eqnarray}
where $\rho(m)$ is the mass spectrum of hadrons,  $\sigma(E)$ is the density of states for hadronic systems, and $a$ is given by
\begin{eqnarray}
a=\frac{\gamma V_o}{2\pi^2 \beta_o^ {3/2}}\,,
\end{eqnarray} 
with $\gamma$ and $b$ being constants and $V_o$ being the interaction volume. Here $e_q^x$ is the q-exponential function given by
\begin{equation}
 e_q^x=[1+(q-1)\,x]^{\frac{1}{q-1}}\,.
\end{equation}

An important consequence of the self-consistency principle is the existence of a limiting effective temperature, $T_o=1/\beta_o$, and of a limiting entropic parameter, $q_o$.

In this work we perform a systematic analysis of experimental data on $p+p$ collisions to verify if there exist $T_o$ and $q_o$ that allow a correct description of all experimental data. The set of experiments used in the present analysis is summarized in Table \ref{tab:Expe}, and covers a wide energy range.

    \begin{table}[!h]
    \centering
    \caption{Set of experimental data for $p+p$ collisions.}
    \begin{tabular}{cccc}\toprule
    Experiment                             &     Energy (GeV)              &	$|\eta|$     \\      \midrule
    CMS (LHC)                               &     7000 \cite{CMS7TeV}           &	$|\eta| < 2,4$     \\
    CMS (LHC)                               &     2360 \cite{CMS7TeV}           &	$|\eta| < 2,4$     \\
    CMS (LHC)                               &     900  \cite{CMS7TeV}           &	$|\eta| < 2,4$     \\
    ALICE (LHC)                             &     900  \cite{ALICE900GeV}       &	$|\eta| < 0,8$     \\
    ATLAS (LHC)                             &     900  \cite{ATLAS900GeV}       &	$|\eta| < 2,5$     \\
    RHIC (BNL)                              &     200  \cite{RHIC-STAR-AuAu}    &	$ 3,3 < \eta < 5,0$     \\       \bottomrule
    \end{tabular}
    \label{tab:Expe}
    \end{table}


\section{Transverse momentum distribution in non-extensive statistics}

A direct method to obtain the effective temperature of the hadronic system produced in hadron-hadron collisions is through the study of the transverse momentum ($p_T$) distribution of secondaries. According to the non-extensive formalism proposed in Ref.~\cite{Bediaga}, the $p_T$-distribution is given by
\begin{equation}
 \frac{1}{\sigma}\frac{d\sigma}{d p_T}=c p_T \int_0^{\infty} \bigg[1+(q-1)\beta \sqrt{p_L^2+p_T^2+\mu^2}\bigg]^{-q/(q-1)}\, dp_L\,,
\end{equation}
where $p_L$ is the longitudinal momentum and $\mu$ is the mass of the hadron.

   \begin{figure}[!h]
       \centering
                   {\label{fig:T_H-T-p+p}\includegraphics[width=10.cm]{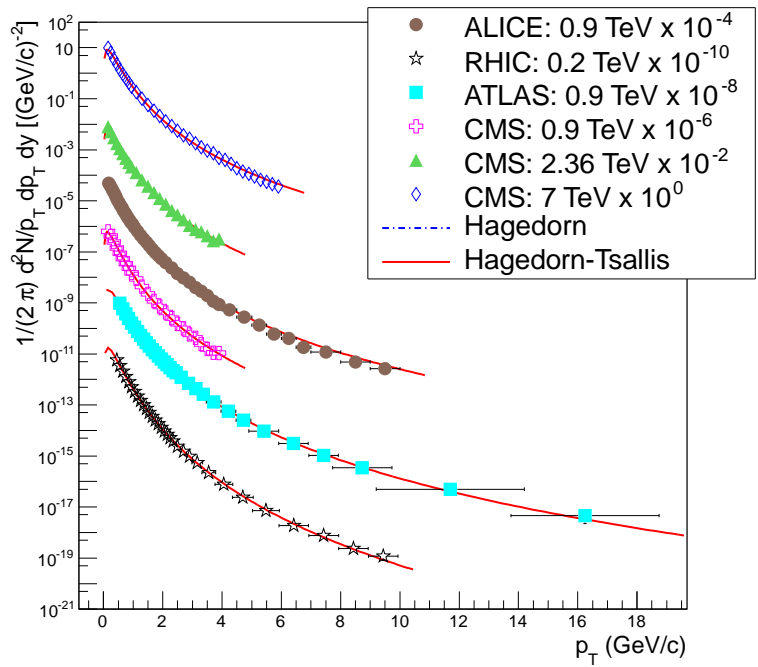}}
          \caption{Fittings of Eq.~\ref{pt} to experimental $p_T$-distributions.}
       \label{ppfit}
   \end{figure}

An useful approximation was proposed in Ref.~\cite{Beck} assuming that $\mu/p_L \approx 0$, resulting
\begin{eqnarray}
\frac{1}{\sigma}\frac{d\sigma}{d p_T}=c[2(q-1)]^{-1/2}B\bigg(\frac{1}{2},\frac{q}{q-1}-\frac{1}{2}\bigg)u^{3/2}[1+(q-1)u]^{-\frac{q}{q-1}+\frac{1}{2}}\,,
\label{pt}
\end{eqnarray}
where $u=\beta p_T$, and $B(x,y)$ is the Beta-function. The dependence of the distribution on the entropic factor, $q$, and on the temperature, given by $\beta = 1/T$, enable us to obtain both parameters by fitting the above expression to experimental data. This procedure has been already used by many authors~\cite{Bediaga, Beck, CMS7TeV, ALICE900GeV, ATLAS900GeV, RHIC-STAR-AuAu}.
It is important to notice that Equation~\ref{pt} should be regarded as an approximation to the distribution one would obtain from the Tsallis entropy by using the thermodynamic relations~\cite{Deppman,Cleyman_Worku}, but to the use made here the approximation is good enough. 

Concerning the approximation on the mass, with the assumption that $\mu/p_L \approx 0$, it should be mentioned that it can be avoided~\cite{Cleyman_Worku} with the restriction that one can not apply the formula obtained for general hadron ($h^+$ and $h^-$), but only on specific particle distributions. Since in this work we aim to make a systematic analysis, and due to the larger number of information on ($h^++h⁻$)-distributions,  we opt to use the approximated formula~\ref{pt}. Even being a good approximation, it causes the effective temperature obtained to be slightly shifted to higher values. This effect can be easily observed in Figure 5 of Ref.~\cite{Cleyman_Worku}, where it can be seen that the peaks of the $p_T$-distributions for heavier particles are shifted to higher values with respect to that for pions.

\begin{figure}[!h]
       \centering
       \subfigure[]
                   {\label{fig:T_H-T-p+p}\includegraphics[width=7.cm]{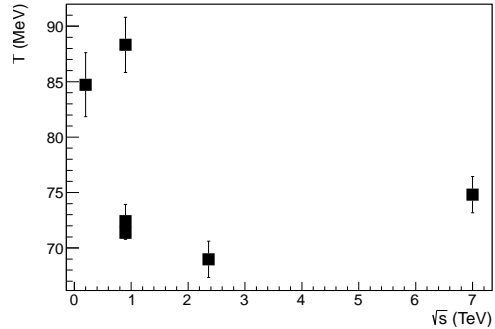}}
       \subfigure[]
                   {\label{fig:q_H-T-p+p}\includegraphics[width=7.cm]{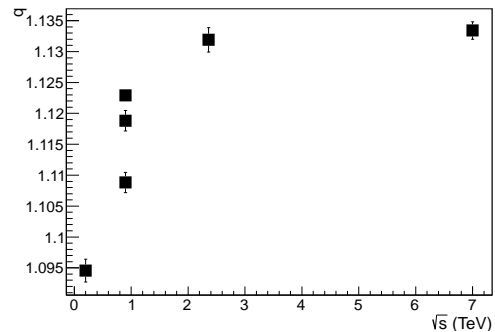}}
          \caption{The values for the parameters corresponding to the best fitted curve to $p_T$-distributions. (a) $T$ as a function of center-of-mass energy. (b) $q$ as a function of center-of-mass energy.}
       \label{fig:p+p_qANDt_H-T_-TodosPt}
   \end{figure}

Applying the method described above to $p+p$ collisions we get nice fittings  to the experimental $p_T$-distributions, as shown in Figure~\ref{ppfit}. We observe  that  Eq.~\ref{pt} correctly describes all data for $p_T$ up to 18~GeV/c.  From these fittings we obtain the results for $T$ and $q$ that are shown in Fig~\ref{fig:p+p_qANDt_H-T_-TodosPt}. The temperature varies inside a relatively narrow range between 70~MeV and 90~MeV for all collisions with center-of-mass energy from 0.2 TeV up to 7 TeV. For energies above 0.9~TeV the temperature can be considered constant with $T\sim$ 73~MeV. Also in the case of the parameter $q$ the variations are relatively small in the energy range studied, and above 0.9~TeV it is approximately constant with $q\sim$~1.13.

\begin{figure*}[htp]
          \centering
          \subfigure[RHIC: 200 GeV]
                   {\includegraphics[width=6.55cm]{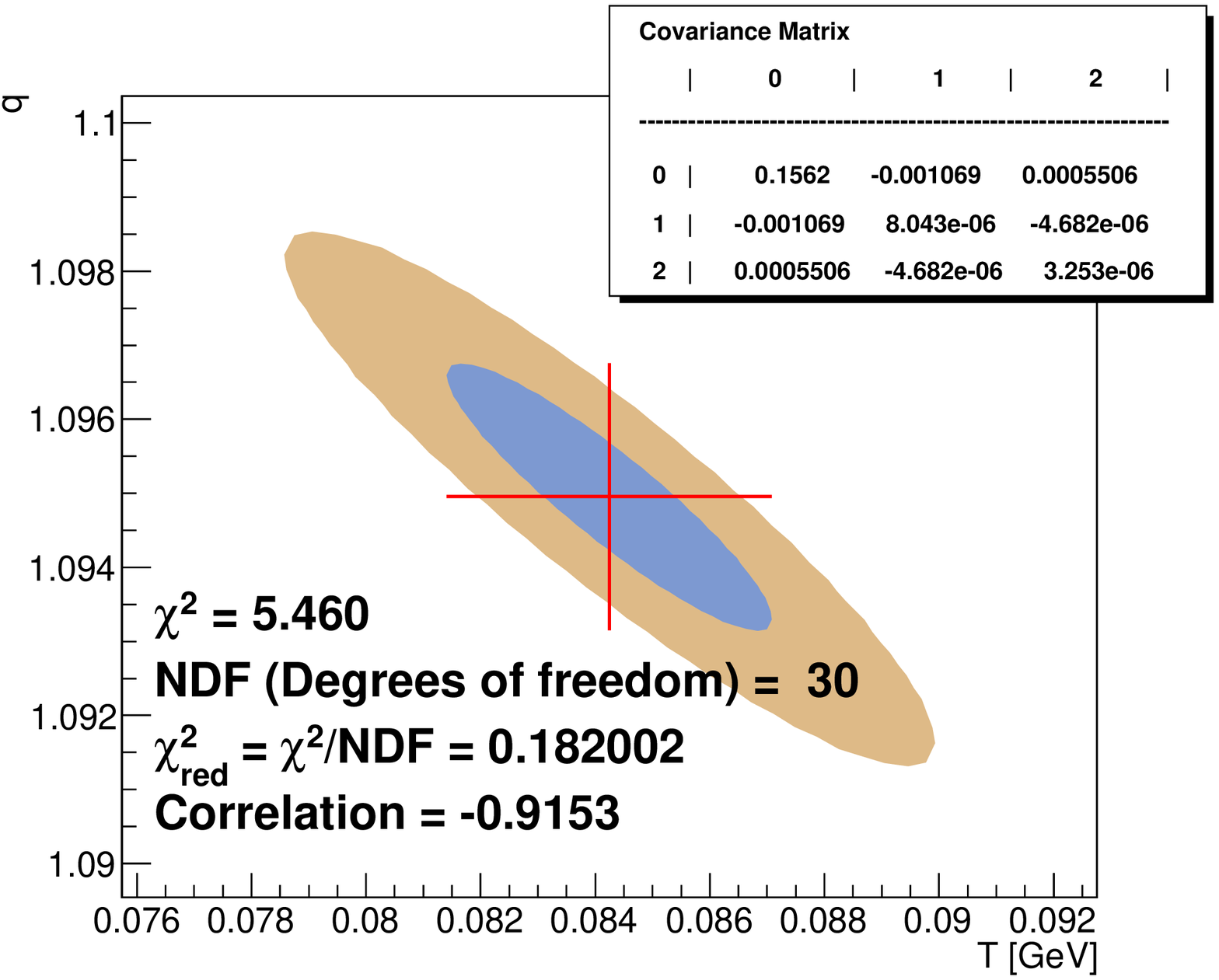}}
          \subfigure[ALICE: 900 GeV]
                   {\includegraphics[width=6.55cm]{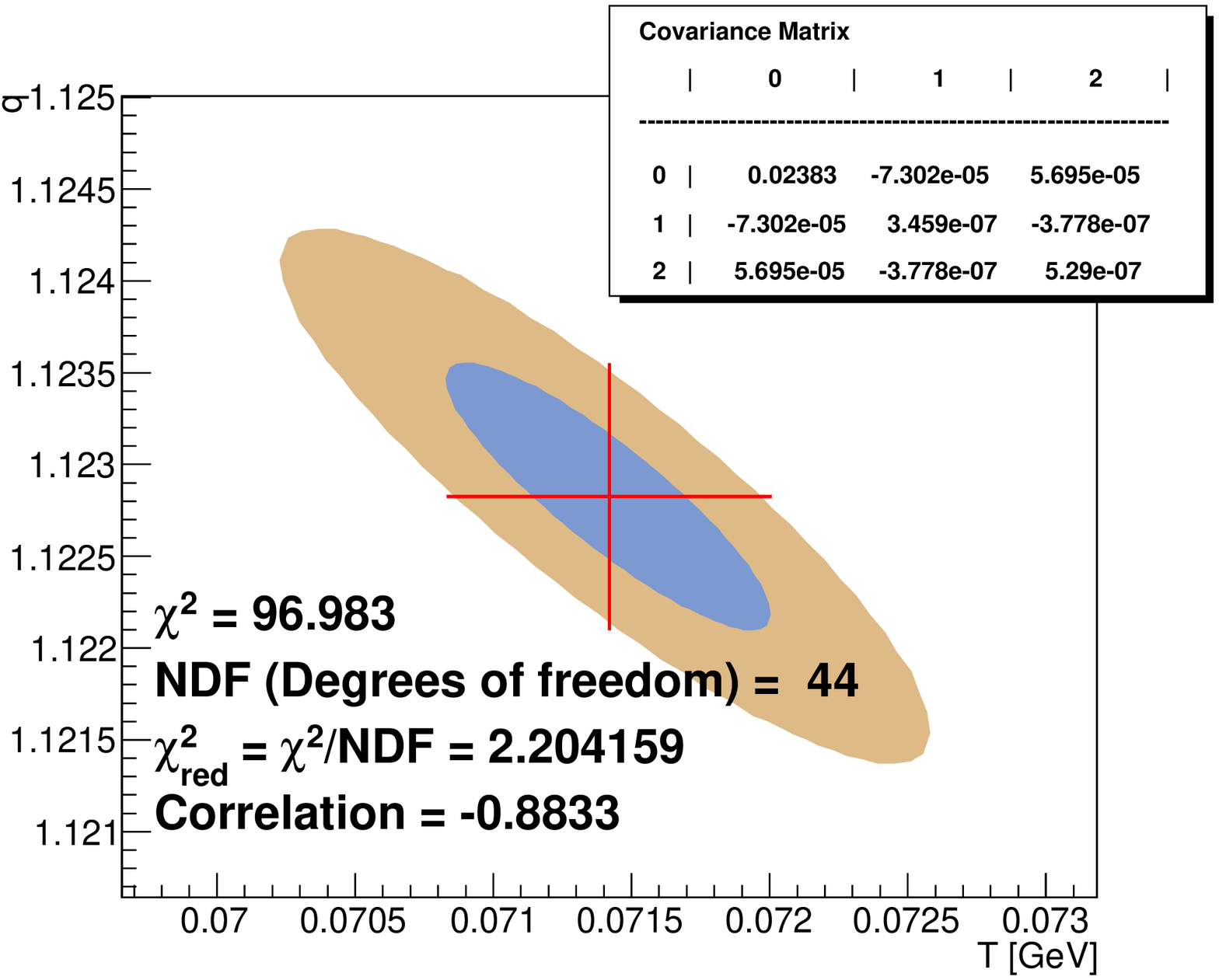}}

          \quad
          \subfigure[ATLAS: 900 GeV]
                   {\includegraphics[width=6.55cm]{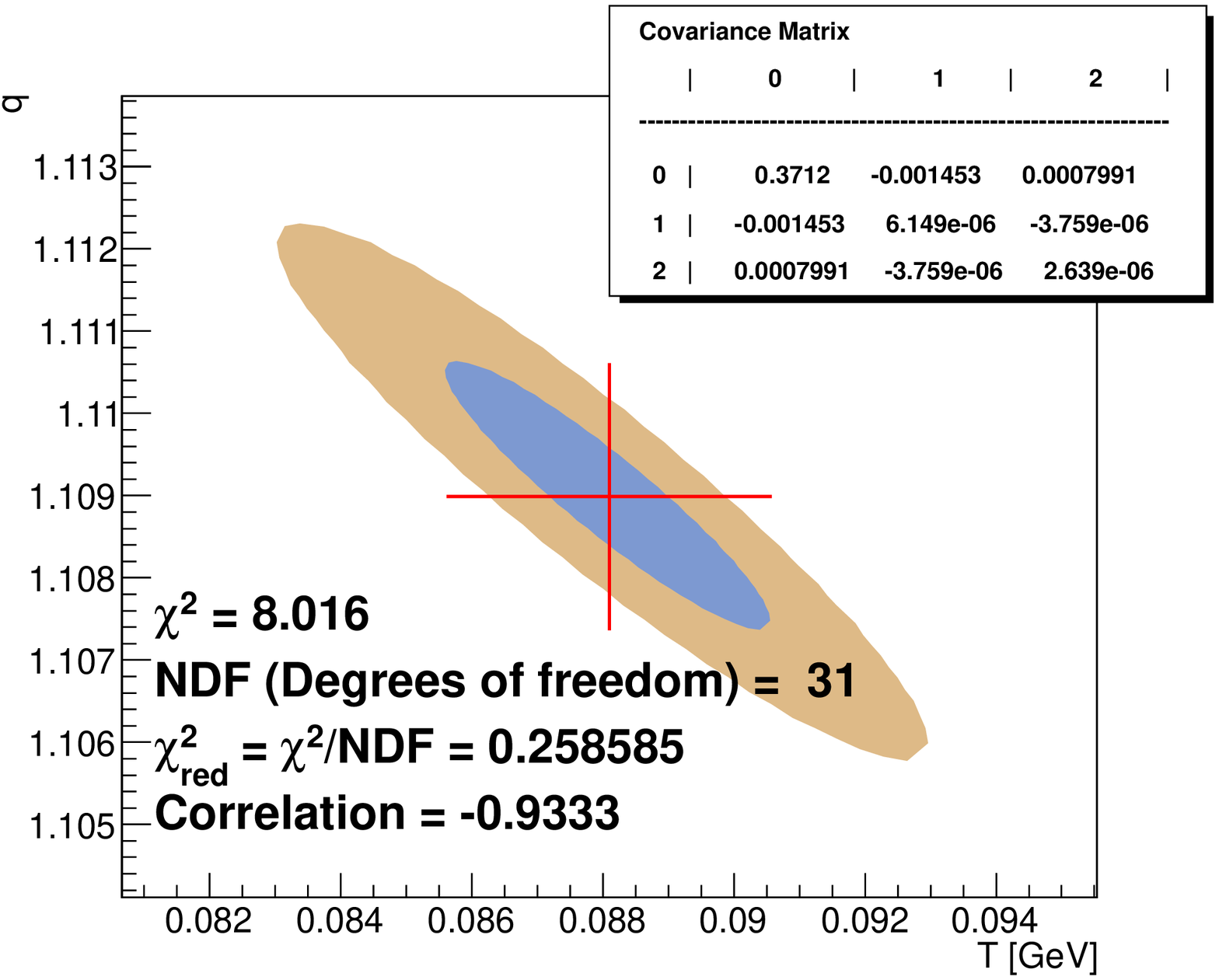}}
          \subfigure[CMS: 900 GeV]
                   {\includegraphics[width=6.55cm]{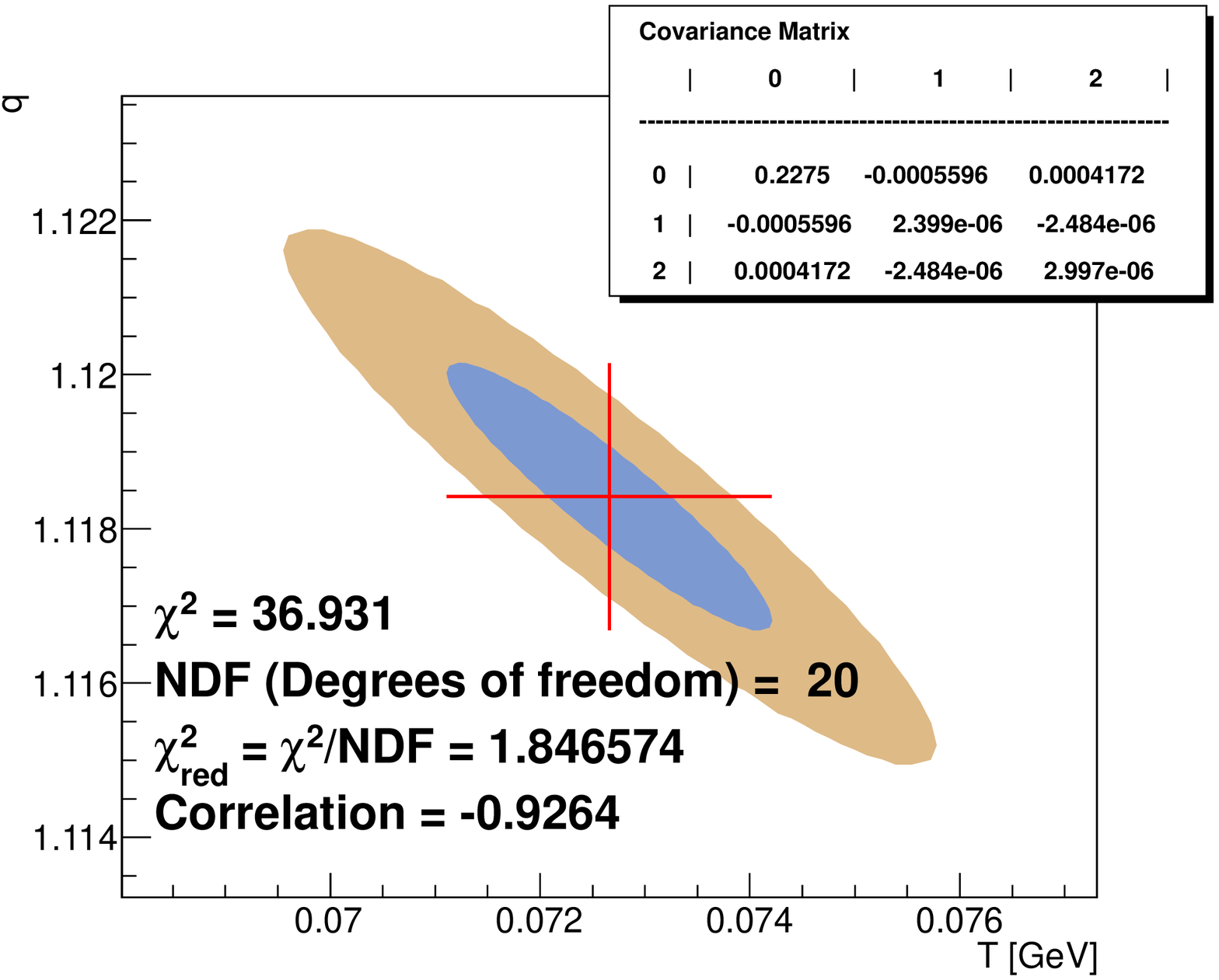}}

          \quad
          \subfigure[CMS: 2.36 TeV]
                   {\includegraphics[width=6.55cm]{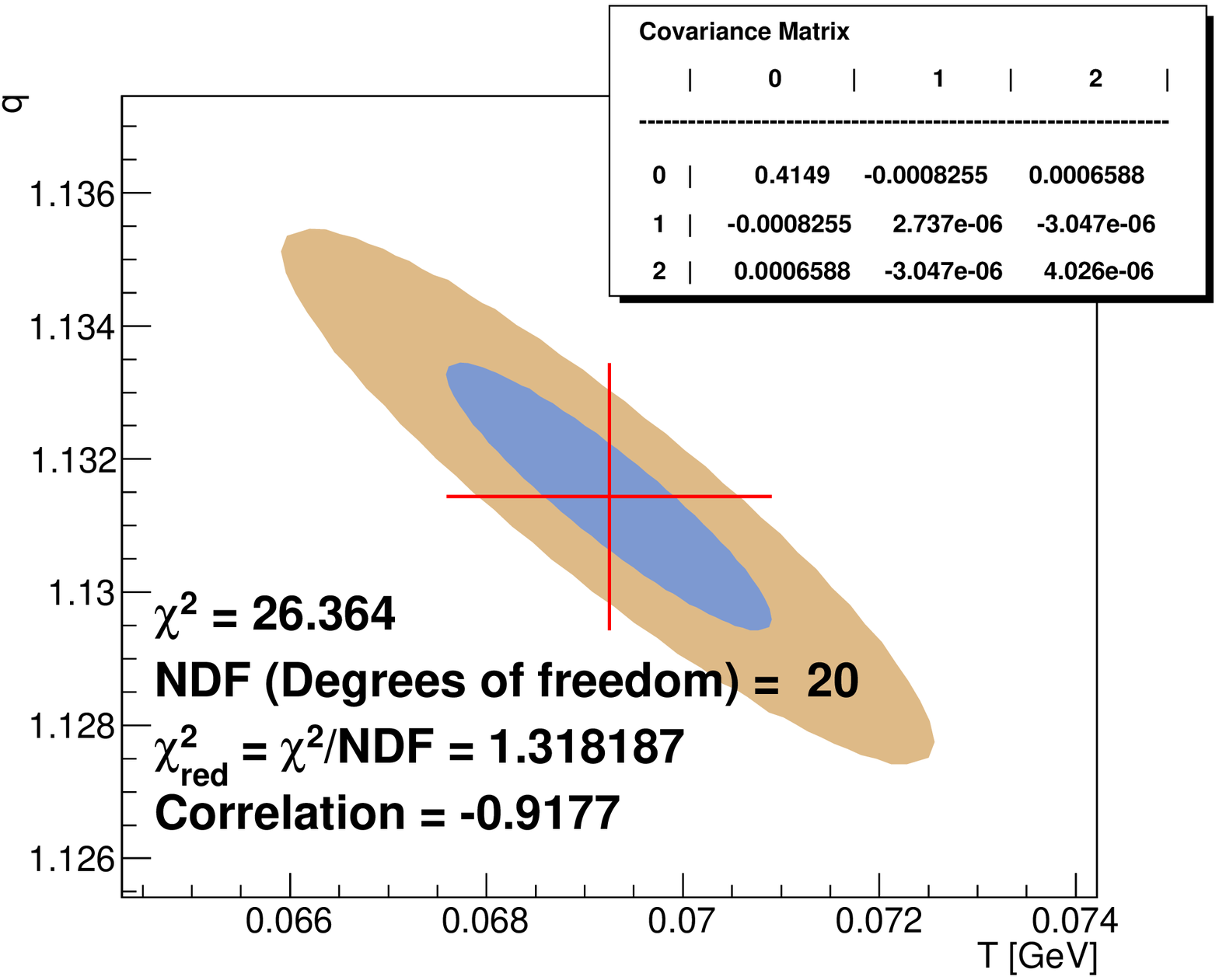}}
          \subfigure[CMS: 7 TeV]
                   {\includegraphics[width=6.55cm]{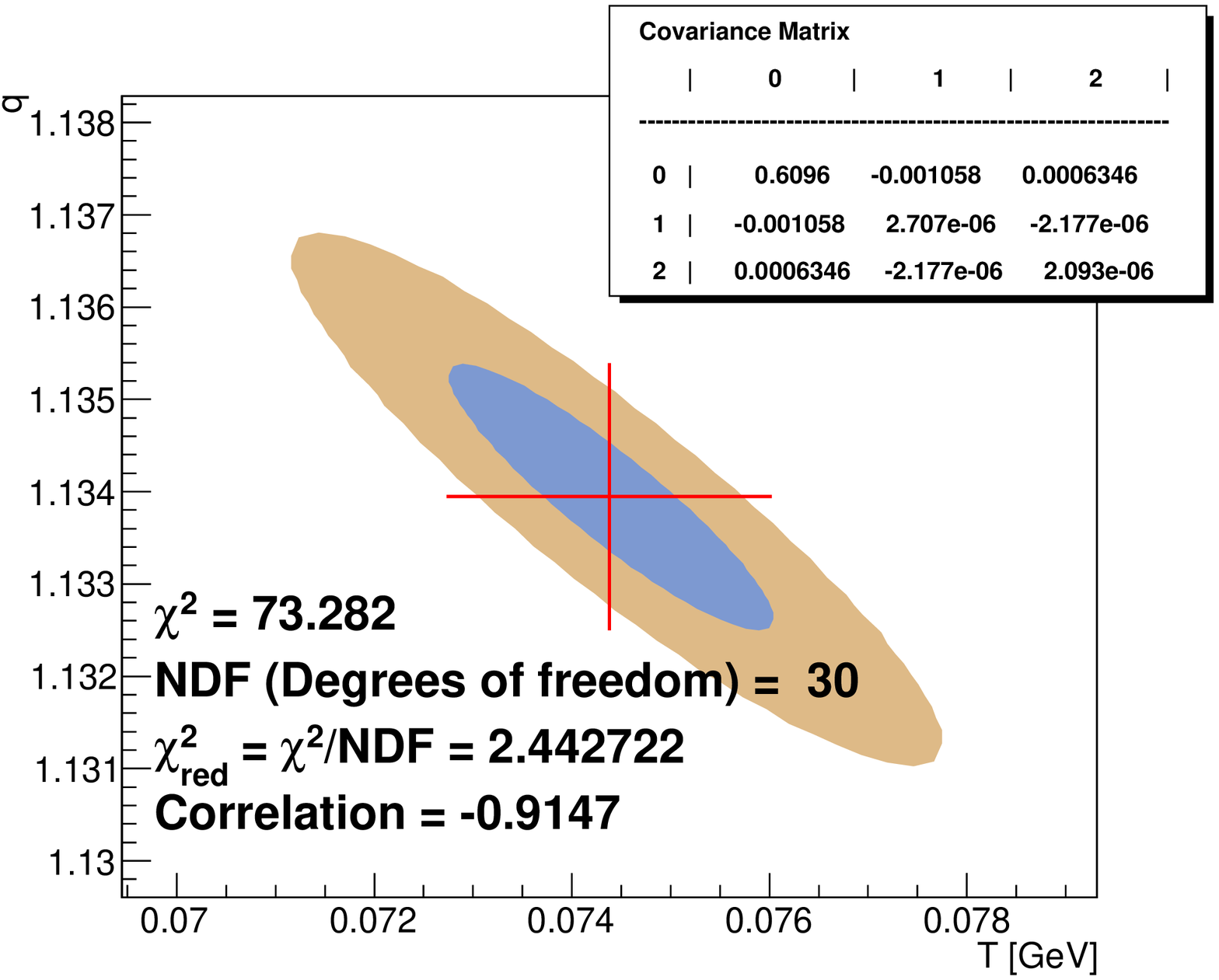}}

          \caption{Analysis of the correlation between the parameters $T$ and $q$. $\chi^2$ is shown as a function of $T$ and $q$  and the cross show the position of the best fit values for the parameters.}
      \label{fig:Cu+Cu_200_TodosPt_0-6-Estatistica}
      \end{figure*}

It is important to notice that $T$ and $q$ in Eq.~\ref{pt} are not completely independent. There is a strong correlation between the two parameters, as can be observed in Fig.~\ref{fig:Cu+Cu_200_TodosPt_0-6-Estatistica} where we show the 2D-plot of  $\chi^2$ distribution for the fittings shown in Fig.~\ref{ppfit}. We clearly see the ellipses that evidences the correlation between the two parameters. Due to their correlation, as we vary $T$ and $q$ simultaneously along the line corresponding to the major axis of the ellipse, the $\chi^2$ remains practically unchanged. This means that it is possible to obtain good $\chi^2$ with pairs ($T$, $q$) near the optimum point found in the fitting, indicated by  crosses in Figure~\ref{fig:Cu+Cu_200_TodosPt_0-6-Estatistica}.

A linear behaviour between $T$ and $q$ was already predicted by Wilk and Wlodarczyk~\cite{Wilk2007,Wilk2009}, who proposed the relation
\begin{equation}
 T=T_o+(q-1)\,c\,,
\label{Tq}
\end{equation}
where $c$ is a constant depending on the energy transfer between the source and its surroundings and on thermodynamical properties of the medium~\cite{Wilk2012}. Since $T \rightarrow T_o$ as $q \rightarrow 1$, $T_o$ is considered to be the Hagedorn's temperature.

Assuming that Eq.~\ref{Tq} establishes a causal relation between $T$ and $q$, then correlation is a necessary consequence~\cite{Rodgers}. Also, it is possible to obtain the regression coeficient, $c$, from the correlation. From the ellipses observed in Fig.~\ref{fig:Cu+Cu_200_TodosPt_0-6-Estatistica} we obtain $T_o=$(192$\pm$15)~MeV and $c=$-(0.95$\pm$0.10)~GeV. It is interesting to notice that the value for $T_o$ found here is in agreement with the critical temperature obtained in lattice QCD calculations~\cite{Ejiri, Miura}.


\section{Testing the constant $T$ hypothesis}

\begin{figure}[h]

          \centering
                      {\includegraphics[width=10cm]{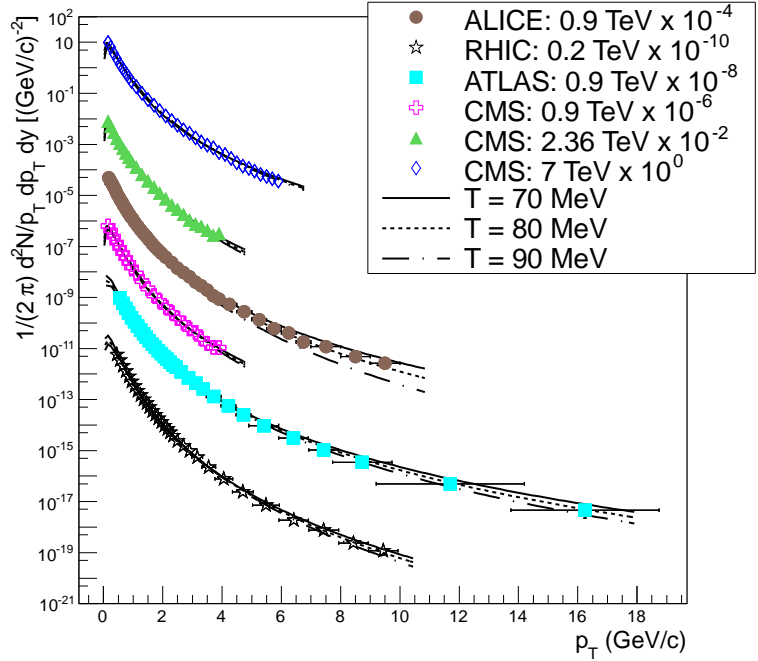}}
   \caption{Typical results for fitting of Eq.~\ref{pt} to experimental data with only $q$ as a free parameter.  The calculations are performed for three different temperatures.}
          \label{fig:Au_62.4_Tconst}
      \end{figure}

Now we slightly modify the fitting procedure by adopting a fixed value for $T$, as suggested by the self-consistency principle~\cite{Deppman} and by the results shown in Fig.~\ref{fig:p+p_qANDt_H-T_-TodosPt}. We used different values for $T$ from 60 MeV up to 120 MeV. Some results are shown in Fig.~\ref{fig:Au_62.4_Tconst} and we observe good fittings of Eq.~\ref{pt} to the data in all cases, now with only $q$ being a free parameter.


These results show that we can fit all experimental data for $p+p$, for $\sqrt{s}$ ranging from 0.2 TeV up to 7 TeV with a fixed temperature, $T$, and using only $q$ as free parameter. Also, the new procedure allows us to study $q$ as a function of $\sqrt{s}$ for different values of $T$. 

\begin{figure}[!h]
       \centering
                   {\label{fig:AllDatas_qSeveralT-a}\includegraphics[width=10.0cm]{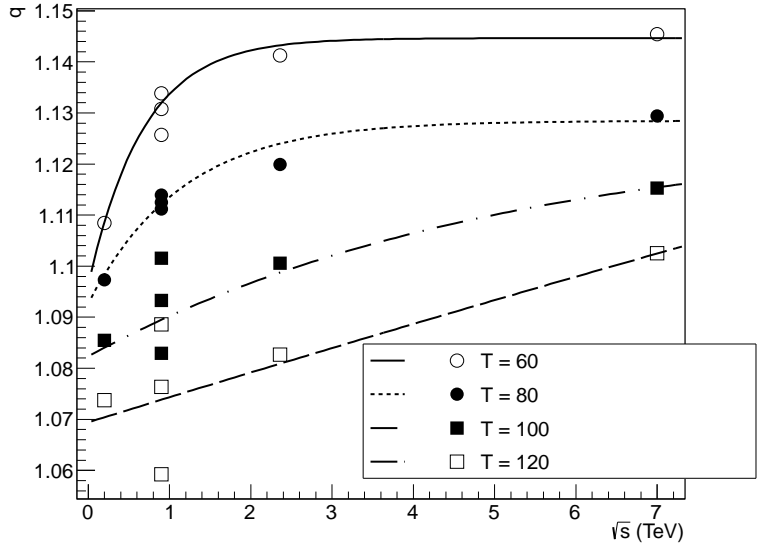}}
          \caption{The values for $q$ corresponding to the best fit with fixed $T$, for different values of temperature. Lines represent the best fitted curves for a sigmoidal function to the data.}
       \label{fig:AllDatas_qSeveralT}
      \end{figure}

In Fig.~\ref{fig:AllDatas_qSeveralT} we show $q$ {\it vs} $\sqrt{s}$ as obtained for different values of $T$. We observe that $q$ increases monotonically with $\sqrt{s}$, the shape being approximately described by a sigmoidal function of the temperature. A sigmoidal behaviour was already conjectured in Ref.~\cite{Beck}, but more experimental information is needed before drawing any canclusion, mainly at low energies. 

From the results shown in Fig.~\ref{fig:Au_62.4_Tconst} we see that the best fittings are obtained for $T\approx$~80~MeV, and in Fig.~\ref{fig:AllDatas_qSeveralT} we observe that for $T=$~80~MeV the entropic paramenter is approximately constant from $\sqrt{s}=$~1GeV to 7~GeV, with $q\approx$~1.12. 

The results obtained here are in agreement with an analysis performed in~\cite{Cleyman_Worku}, where the $p_T$-distributions for different particles produced in $p+p$ collisions at center-of-mass energy of 0.9 TeV were studied, resulting in $T$ and $q$ constant for all particles with $T\approx$~75 MeV and $q\approx$~1.15. There are also indications that these results hold for $A+A$ collisions~\cite{Sena}.

\section{Conclusions}

In this work we present a systematic analysis of $p_T$-distributions observed in $p+p$ experiments. It is shown that the experimental data gives support to the hypothesis of a limiting effective temperature, $T$, and a limiting entropic factor, $q$.

In the analysis, we show that $T$ and $q$ are correlated parameters in the fitting procedure of the theoretical transverse momentum distribution to the experimental data. From this correlation, it results that the critical temperature is $T_o=$(192$\pm$15)~MeV, a value in good agreement with lattice-QCD predictions. 

The study presented here gives  evidences for a limiting effective temperature for the hadronic system formed in hadron-hadron collisions, with $T\approx$~80~MeV, and also of a limiting entropic factor with $q\approx$~1.12. These results are in good agreement with values found in Literature.

\section{Acknowledgements}

The authors thanks Prof. Dr Otaviano Helene (Universidade de São Paulo) for interesting discussions about the statistical analysis of the experimental data presented here.
This work received support from the Brazilian agency, CNPq, under grants 305639/2010-2 (A.D.) and 136988/2009-1 (I.S.).

\bibliographystyle{aipproc}

\end{document}